\documentclass[aps,eqsecnum,pra,twocolumn]{revtex4}
\usepackage{graphics,graphicx}
\usepackage{amsmath}
\usepackage{amssymb,latexsym,mathrsfs}
\usepackage{hyperref}

\def\bea{\begin{eqnarray}}
\def\eea{\end{eqnarray}}
\def\ba{\begin{array}}
\def\ea{\end{array}}

\def\beq{\begin{equation}}
\def\eeq{\end{equation}}

\begin{document}

\title{Quantum Path predictability for an electronic Mach-Zehnder interferometer in presence of environment induced Decoherence and Quantum Erasing process}

\author{Samyadeb Bhattacharya$^{1}$ \footnote{sbh.phys@gmail.com}, Sisir Roy$^{2} $ \footnote{sisir.sisirroy@gmail.com}}
\affiliation{$^{1}$Physics and Applied Mathematics Unit, Indian Statistical Institute, 203 B.T. Road, Kolkata 700 108, India \\
	$^{2}$National Institute of Advanced Studies
	IISC Campus , Bangalore 56 0012
	India.\\}

\vspace{2cm}
\begin{abstract}

\vspace{1cm}

\noindent In this paper, we have estimated the temperature dependent path predictability for an electronic Mach-Zehnder interferometer. The increment of path predictability can directly be associated with stronger decoherence process. We have also theoretically predicted that placing two detectors in both the paths, which are at the same equilibrium temperature with the system, erases all the memory of path information and hence acts like a quantum eraser. 

\vspace{2cm}


\end{abstract}

\vspace{1cm}

\maketitle

\section{Introduction}

Quantum superposition is one of the elementary properties that distinguishes quantum mechanics from it's classical counterpart. This specific property of quantum states gives rise to the phenomena of coherence, which though very much familiar in classical wave theory, appears in an apparently mysterious way since the effect arises from the summation of probability amplitudes rather than the physical field amplitudes. The amount of interference that can be observed in a quantum interference experiment, depends on the anonymity of the path that is taken by the particle whose interference property is under observation. The process of gaining information about the path is complimentary to the process of producing interference pattern. Because gaining information about the path of the particle involves measurement processes, which destroys the superposition of the available states making the behavior of the particle classically probabilistic. If we discuss this issue in a more quantitative manner, then we must firstly specify a common measure giving the amount of interference, known as fringe visibility $V$, given by

\beq\label{1.1}
V=\frac{I_{max}-I_{min}}{I_{max}+I_{min}}
\eeq

where $I_{max}$ and $I_{min}$ are the observed maximum and minimum intensity respectively. From the expression given in \ref{1.1} we can clearly see that the visibility measures the amount of contrast in interference fringe pattern. For maximally coherent beam the visibility is equal to unity and with the decay of coherence it decreases, reducing the contrast of the interference fringes. The concept of complementarity, which is one of the most fundamental notion of quantum mechanics, can be brought into consideration in this respect. A few years back, Englert \cite{englert} has shown that there exists a fundamental inequality between the measured distinguishibility and visibility for a Welcher Weg  type measurement. Distinguishibility represents the quantification of `which way information' in a quantum interference experiment. If the interfering quantum entity is entangled with a different quantum system, then it's path becomes somewhat marked and hence the anonymity of the path vanishes along with the interference pattern in the detector. Englert, in his paper \cite{englert} showed that, the derivation of the complimentary relation between distinguishibility and visibility is logically independent of any uncertainty principle and relies solely on unitary transformation of the system and the entangled meter state. Englert's analysis can also be included in the discussion of Quantum erasure \cite{scully1,scully2,Kwiat}. Quantum erasure is a particular technique to protect the interference fringe pattern in an interference experiment, at the expanse of the respective complimentary knowledge of which way information. This is fundamentally different from any kind of simple detector noise, which though destroys the interference pattern, but does not trade it for complimentary information.\\
In this paper, we will consider the which way information experimental situation from the backdrop of dissipative interaction. By dissipative interaction, here we mean the coupling of the concerning system with the meter state, which is, essentially considered to be a thermal magnetic field. We will study the effect of decoherence induced by the thermal magnetic field. Here the system under consideration is an electronic analogue of Mach - Zehnder interferometer as presented by Roulleau et.al. \cite{roul}. Before going into the detail of our work, we must present a short discussion of the experiment, based on which we are constructing our theoretical work. For detail of the experiment, the authors refer to Roulleau et.al. \cite{roul}. Constructing an electronic device that mimics the optical Mach-Zehnder interferometer, is a remarkable achievement in the opto-electronic branch of experimental physics. Ji et.al \cite{Ji} has constructed such kind of a device operating in the Quantum Hall regime, where the transport occurs through one dimensional chiral channels. Those channels perfectly mimic the photon beams of optical Mach-Zehnder interferometer. To mimic the effect of decoherence, Roulleau et.al. introduced a voltage probe in the path of the electron, which essentially destroys the phase memory of the electron. The inclusion of the voltage probe in the path of the electron, removes the ambiguity of the trajectory of the particle and which is followed by the suppression of interference effect. This process has been used by other authors for Mach-Zehnder interferometer to predict current fluctuations in presence of decoherence. Roulleau et.al. used a voltage probe to introduce energy redistribution, for an electronic Mach-Zehnder interferometer acting in the quantum Hall regime. The probe is realized with a quantum point contact in one arm of the interferometer. A quantum point contact is nothing but a  tunnel barrier of adjustable height for electrons near Fermi level. This property has been frequently used by many physicists to inject and detect electrons in a confined region of electron gas. A perpendicular magnetic field can be used in the point contact region to modify the nature of the quantum states. In a wide two dimensional electron gas, a perpendicular magnetic field eliminates the two degrees of freedom to form dispersionless Landau levels. Near the boundary, the Landau levels are transformed into magnetic edge channels, capable of freely moving along the boundary. These edge channels are similar to one dimensional sub bands for purely electrostatic lateral confinement in a channel \cite{houten}. In our work, we will consider that the electron passing through quantum point contact (which is acting as a path detector) will interact with a thermal magnetic field and as a result we will lose the visibility of the interference pattern. Based on this experimental backdrop, we will calculate the path predictability of the electron interacting with the thermal magnetic field and show how it varies with temperature. Based on this we will discuss about the issue of quantum erasure and also simultaneous measurement of complimentary informations on the backdrop of open quantum systems. The reason behind the study of path predictability and visibility as a function of temperature is to study their complementary relationship with increasing or decreasing environmental interactions. Firstly we will present the derivation of path predictability for an electronic Mach-Zehnder interferometer with a quantum point contact as a path detector. There we will see how the predictability depends on the decay parameter. After that we will derive the explicit form of the decay parameter in connection to decoherence, for a spin half particle interacting with a thermal magnetic field. As a result, we will find the temperature dependence of the path predictability. In the end we will conclude with some possible implications. 


\section{Temperature dependence of path predictability}

To calculate the path predictability, we will consider the procedure developed by Krause et.al in a recent paper \cite{krause}. As we have mentioned earlier, that to quantify which way information, the path distinguishablity $D$ can be a useful parameter \cite{jaeger,pdd}. For a two path interferometer, there is a relation between distinguishablity and visibility \cite{englert}, given by

\beq\label{2.1}
D^2 + V^2 \leq 1
\eeq  

On the other hand, which way information may also be quantified by another parameter path predictability \cite{englert,jaeger,pdd}, given by the expression

\beq\label{2.2}
P = |p_1-p_2| 
\eeq

where $p_1$ and $p_2$ are the probabilities for the particles to take the paths 1 and 2 respectively. Path predictability also obeys same kind of relation with the visibility \cite{englert,krause,jaeger,pdd} 

\beq\label{2.3}
P^2 + V^2 \leq 1
\eeq

So increment in path predictability naturally results in the loss of visibility of the interference pattern. In other words we can say that with increment of path predictability the system becomes more and more classical. Krause et.al. \cite{krause} has considered the problem of determining the path predictability for an unstable particle in two path Mach-Zehnder type interference experiment set up. They have considered a non-relativistic Schr\"{o}dinger equation describing the center of mass motion of an excited atom given by

\beq\label{2.4}
i\hbar \frac{\partial \psi(\mathbf{r},t)}{\partial t} = \left(mc^2-i\frac{\hbar\Gamma}{2}-\frac{\hbar^2}{2m}\nabla^2\right) \psi(\mathbf{r},t)
\eeq

where $\Gamma$ is the decay rate of the particle. Following Greenberger and Overhauser \cite{green}, Krause et.al. \cite{krause} found the solution of this equation as 

\beq\label{2.5}
\psi(\mathbf{r}) = A \exp \left[ i\frac{\mathbf{p.r}}{\hbar}\left(1+i\frac{m\hbar\Gamma}{2p^2} \right)  \right]
\eeq

Here the the magnitude of the wave vector is considered to be complex and which is given by 

\beq\label{2.6}
|\mathbf{k}|^2 = (k+i\kappa)^2 \simeq \frac{p^2}{\hbar^2} + i\frac{m\Gamma}{\hbar}
\eeq

where $p=\hbar k$ is particle's momentum, $\kappa$ is the imaginary part  and $k$ is the real part of the wave vector respectively.  

\beq\label{2.7}
\kappa\simeq m\Gamma/2p=1/2l
\eeq

So this implies that the probability amplitude for a particle to travel a distance $x$ can be given by 

\beq\label{2.8}
\psi_x = A e^{ipx/\hbar} e^{-x/2l}
\eeq

The second exponential term in the right hand side of equation \ref{2.8} represents dissipation. Here one one has to keep in mind that the energy loss due to dissipation must be very small compared to the initial kinetic energy of the particle, i.e. $\hbar\Gamma \ll p^2/2m$. Now unlike Krause et.al. \cite{krause}, we take that the two paths $\mathbf{R_1}$ and $\mathbf{R_2}$ are independent of any external dissipative interactions. We take this assumption for simplicity. Also $|\mathbf{R_1}|=|\mathbf{R_2}|$, i.e. the length of the two arms of the interferometer are same. Now we state that there is a path detector in the form of a magnetic quantum point contact of length $L_{dec}$ is placed in the path $\mathbf{R_2}$. 

So the wave functions corresponding to the particles taking path $\mathbf{R_1}$ and $\mathbf{R_2}$ are respectively given by \cite{krause}

\beq\label{2.9}
\begin{array}{ll}
	\psi_1=-\frac{i}{2}e^{ipR_1/\hbar}\\
	\psi_2=-\frac{i}{2}e^{ipR_2/\hbar}e^{-L_{dec}/2l}
\end{array}
\eeq 

Now according to Krause et.al. \cite{krause}, the normalized form of path predictability, which is compatible with the definition \ref{2.2}, can be given by

\beq\label{2.10}
P = \left|\frac{|\psi_1|^2-|\psi_2|^2}{|\psi_1|^2+|\psi_2|^2}\right|
\eeq 

Putting the values of $\psi_1$ and $\psi_2$, we get the expression of path predictability as

\beq\label{2.11}
P=\frac{1-e^{-L_{dec}/2l}}{1+e^{-L_{dec}/2l}}=\tanh\left(\frac{mL_{dec}}{4\hbar k}\Gamma\right)
\eeq

Now we want to estimate the explicit expression of the decay parameter $\Gamma$, by considering the interaction of a spin half particle passing through a thermal magnetic field, in the region of quantum point contact.\\
For this purpose, we consider a two level atom with a vacuum thermal magnetic field. In the point contact region, the atom is interacting with the many modes of the vacuum radiation field, which is considered as a collection of harmonic oscillators. The total Hamiltonian of the system, filed and interaction is given by 

\beq\label{2.12}
H_T = H_S + H_R + H_I 
\eeq

where $H_S$, $H_R$ and $H_I$ are the system, reservoir and interaction Hamiltonian respectively \cite{car}. 

\beq\label{2.13}
\begin{array}{ll}
	H_S = \frac{1}{2}\hbar \omega_A \sigma_z\\
	H_R = \sum_{\mathbf{v},\lambda}\hbar\omega_v b^{\dagger}_{\mathbf{v},\lambda} b_{\mathbf{v},\lambda}\\
	H_I = \sum_{\mathbf{v},\lambda} \hbar (\beta^{*}b^{\dagger}_{\mathbf{v},\lambda} \sigma_{-} + \beta b_{\mathbf{v},\lambda}\sigma_{+})
\end{array}
\eeq 

here $\omega_A$ is the atomic frequency and the complex $\beta$ is the interaction strength given by 

\beq\label{2.14}
\beta = -i e^{i\mathbf{v.R_2}} \sqrt{\frac{\omega_v}{2\hbar\epsilon_0 \phi}} \hat{\mathbf{e}}_{\mathbf{v},\lambda}.\mathbf{d}_{21}
\eeq

$v$ and $\lambda$ represents the wave vectors and the polarization states for the reservoir oscillators. $\hat{\mathbf{e}}_{\mathbf{v},\lambda}$ is the unit polarization vector and $\mathbf{d}_{21}$ is the dipole moment vector and $\phi$ is the quantization volume. \\
The master equation for this two level atom interacting with a thermal magnetic field can be stated as \cite{car}

\beq\label{2.15}
\begin{array}{ll}
	\dot{\rho} = -i\frac{1}{2}\omega'_A[\sigma_z, \rho]\\
	~~~~~ +\frac{\gamma}{2}(\bar{n}+1)(2\sigma_{-}\rho\sigma_{+}-\sigma_{+}\sigma_{-}\rho-\rho\sigma_{+}\sigma_{-})\\
	~~~~~
	+\frac{\gamma}{2}\bar{n}(2\sigma_{+}\rho\sigma_{-}-\sigma_{-}\sigma_{+}\rho - \rho\sigma_{-}\sigma_{+})
\end{array}
\eeq 

where $\bar{n}$ is the Planck distribution given by

\beq\label{2.16}
\bar{n} = \frac{1}{\exp(\hbar\omega_A/K_B T)-1}
\eeq

It can be found that 

\beq\label{2.17}
\gamma = \frac{1}{4\pi\epsilon_0} \frac{4\omega^3_A d^2_{12}}{3\hbar c^3}
\eeq

It is the corrected result for Einstein A coefficient \cite{car}.

\beq\label{2.18}
\omega'_A = \omega_A + 2\Delta' + \Delta
\eeq 

where $\Delta'$ and $\Delta$ are corrections due to Lamb shift. 

The solution of \ref{2.15} can be found as 

\beq\label{2.19}
\begin{array}{ll}
	\dot{\rho}_{22} = -\gamma (\bar{n}+1)\rho_{22} + \gamma \bar{n}\rho_{11}\\
	\dot{\rho}_{11} = -\gamma \bar{n} \rho_{11} + \gamma (\bar{n}+1)\rho_{22}\\
	\dot{\rho}_{21} = -\left[\frac{\gamma}{2}(2\bar{n}+1)+i\omega_A\right]\rho_{21}\\
	\dot{\rho}_{12} = -\left[\frac{\gamma}{2}(2\bar{n}+1)-i\omega_A\right]\rho_{12}
	
\end{array}
\eeq

$\rho_{12}$ and $\rho_{21}$ are the cross diagonal components of the density matrix and hence they represents quantum coherence. So their decay, as given by \ref{2.19}, basically represents the process of decoherence of the two state system caused by the interaction with the field. It is very easy then to find the coherence decay parameter from \ref{2.19}, which can be expressed as 

\beq\label{2.20}
\Gamma_{dec} = \frac{\gamma}{2} (2\bar{n} + 1) = \frac{\omega_A^3 d^2_{12}}{6\pi\epsilon_0 \hbar c^3} \coth\left( \frac{\hbar\omega_A}{2K_B T} \right) 
\eeq

Using the this deacy parameter in equation \ref{2.11}, we get the path predictability as

\beq\label{2.21}
P = \tanh \left[\alpha \coth \left(\frac{\hbar \omega_A}{2K_B T} \right)  \right]
\eeq

where

\beq\label{2.22}
\alpha = \frac{mL_{dec} \omega^3_{A} d^2_{12}}{24\pi\epsilon_0 \hbar^2 k c^3}
\eeq

\begin{figure}[htb]
	{\centerline{\includegraphics[width=7cm, height=5cm] {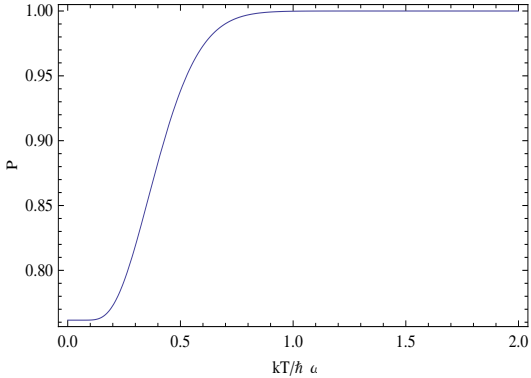}}}
	\caption{$P$ vs. $K_BT/\hbar\omega_A$}
	\label{figVr}
\end{figure}

In FIG. 1, we have shown that how the path predictability varies with increasing temperature. It clearly shows that as the temperature increases, the path predictability also increases and asymptotically reaches it's maximum value 1. This also indicates that, as temperature increases, the visibility of interference fringes decreases to zero and with that the system becomes classically probabilistic. We should also note that the path predictability will reach very close to 1 at and after $K_BT \geq \hbar\omega_A$; i.e. when the kinetic energy part dominates over the quantum energy. Here we also intend to mention that, even at zero or very low temperature, the path predictability has a finite value of 

\beq\label{2.23}
P_{T\rightarrow 0} = \tanh \alpha
\eeq

Here we also want to touch upon the issue of quantum erasing process. erasing is the process by which we can protect the quantum interference pattern in the Welcher Weg type experiment. If by applying some method we can erase the path information of the particle, then we can regain the complimentary information of visibility; hence we get back the contrast of the fringe pattern. Let us now consider that there are two quantum point contact detector of length $L_{dec}$ in both the paths of the Mach-Zehnder interferometer. If the of the two detectors separately in equilibrium with the system be $T_1$ and $T_2$ respectively, then by following the same procedure we find the expression of path predictability to be 

\beq\label{2.24}
P' = \tanh \left[\alpha \left[\coth \left(\frac{\hbar \omega_A}{2K_B T_2} \right) - \coth \left(\frac{\hbar \omega_A}{2K_B T_1} \right)\right]  \right]
\eeq

For simplicity, we have assumed that the $\alpha$s of both the detectors to be the same. Now if the equilibrium temperatures of both the detectors are maintained to be the same, i.e. $T_2=T_1$, then we will find that the modified path predictability $P'=0$. Which means that, if we can maintain the temperatures of the detectors in both the paths of the interferometers to be the same, then we will loose all the information about the path. Hence the complimentary information of the fringe visibility will be regained. To analyze the issue in a more general way, we take the temperature difference between the two detectors to be $\Theta = T_2-T_1$. Let us also consider a situation close to the thermal equilibrium, when $\Theta \ll T_1$. Under this approximation, we find the path predictability to be 

\beq\label{2.25}
	P''= \\
	\tanh \left[\frac{\alpha\hbar\omega_A\Theta}{2K_BT_1^2}\frac{\mathrm{cosech}^2\left(\frac{\hbar\omega_A}{2K_BT_1}\right)}{\left(1-\frac{\hbar\omega_A\Theta}{2K_BT_1^2}\coth(\frac{\hbar\omega_A}{2K_BT_1}) \right) } \right]
\eeq

\begin{figure}[htb]
	{\centerline{\includegraphics[width=7cm, height=5cm] {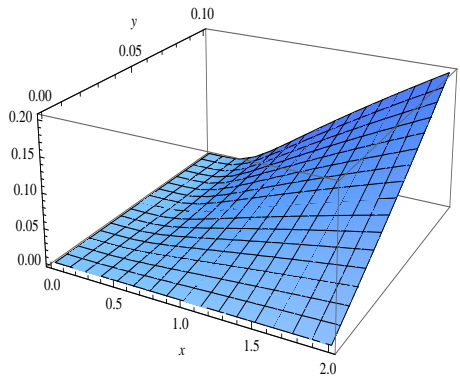}}}
	\caption{$P$ vs. $x=K_BT_1/\hbar\omega_A$ and $y=\Theta/T_1$}
	\label{figVr}
\end{figure}

In FIG 2, we have shown that how path predictability varies with $T_1$ and $\Theta$. The plain region is where path predictability is minimum, i.e. tending to zero. That is when the temperature $T_1$ and (or) the temperature difference between the two detectors $\Theta$ are very small. We can also see that even for high temperature, ($K_BT>\hbar\omega_A$), if the temperature difference between the two detectors is kept very small, the path predictability tends to zero. So our calculation shows that $\Theta$ acts a kind of tuning parameter which maintains the ``quantumness" of the electronic system under study in this kind of experiment. In other words, we can say that the temperature difference between the two detectors acts like a quantum erasure in the electronic Mach-Zehnder experiment.   


\section{Conclusion}

In this paper, we have estimated the path predictability for a which way information experimental set up of an electronic Mach-Zehnder interferometer, in presence of decoherence. The result shows that, in presence of decoherence induced by the thermal magnetic field, the path predictability depends on the temperature of the system and the coupled magnetic environment. We must also mention here that we are considering the system is in thermal equilibrium with the environment. FIG. 1 shows how the path predictability varies with increasing temperature. It shows that the predictability increases quickly as temperature increases and reaches the maximum value of 1. This is an expected outcome in an experimental scenario. Because we can easily assert that, with the increment of thermal energy, the environmental interaction with the system will also increase and with that the process of decoherence should also become stronger. Thermal noise will enhance the process of decoherence. So the situation will then become more and more classical and hence the path predictability will increase and asymptotically reach it's highest value; which indicates the behavior of the system will then be classically probabilistic. We can also see from equation \ref{2.23} that even at zero temperature, there exist a finite path predictability of the value $tanh\alpha$. This is due to the fact that even at zero temperature, the environment does not seize to interact with the system; though the interaction is then considerably weaker than at higher temperature. We can even make it more weak by controlling the tunable parameters like $L_{dec}$, $\omega_A$ and $d_{12}$. Here we have also presented a theoretical prediction that placing two detectors, which have the same equilibrium temperature with the system, in both the paths of the interferometer erases the memory of path information completely. Here the temperature difference between the two detectors acts like a quantum eraser. If this temperature difference can be adjusted to be very small, then even with path detectors having considerable thermal noise we can maintain the quantum coherence of the system.  

\section{Acknoledgement}

Samyadeb Bhattacharya thanks Prof. Guruprasad Kar of Indian Statistical Institute and Prof. Sibasish Ghosh of Institute of Mathematical Science, Chennai for useful discussions. Author also thanks Mr. Souvik Pramanik of Indian Statistical Institute for some technical helps.

\end{document}